\documentclass[aps,pre,longbibliography,twocolumn,floatfix]{revtex4-1}
\usepackage{graphicx}
\usepackage{amsmath}
\usepackage{amssymb}
\usepackage{bm}
\usepackage{times}
\usepackage[pdftex]{hyperref}
\hypersetup{colorlinks=true,linkcolor=blue,citecolor=blue,urlcolor=blue}
\usepackage{verbatim}

\begin{document}

\title{Free energy barriers in the Sherrington-Kirkpatrick model}

\author{T.~Aspelmeier}
\affiliation{Institute for Theoretical Physics,  Georg-August-Universit\"at G\"ottingen,  D37077, G\"ottingen,  Germany }

\author{M.~A.~Moore}
\affiliation{Department  of
Physics and Astronomy, University of Manchester, Manchester M13 9PL, United Kingdom}
\date{\today}

\begin{abstract}

The free energy landscape of the Sherrington-Kirkpatrik (SK)  Ising spin glass is simple in the framework of the Thouless-Anderson-Palmer (TAP) equations as each solution (which are minima of the free energy) has associated with it a nearby index-one saddle point. The free energy barrier to escape the minimum is just the difference between the saddle point free energy and that at its associated  minimum. This difference is calculated for the states with free energies $f > f_c$. It is very small for these states, decreasing as $1/N^2$, where $N$ is the number of spins in the system. These states are not marginally stable. We argue that such small barriers are why numerical studies  never find these states when $N$ is large.  Instead the states which are found are those which have marginal stability. For them the barriers are at least of $O(1)$. $f_c$ is the free energy per spin below  which the states develop broken replica-symmetry like overlaps with each other. In the regime $f < f_c$ we can only offer some possibilities based around  scaling arguments.  One of these suggest that the barriers might become as large as $N^{1/3}$. That might be consistent with recent numerical studies on the Viana-Bray model, which were at variance with the expectations of Cugliandolo and Kurchan for the SK model.
\end{abstract}
\maketitle

\section{Introduction}

The free energy landscape of disordered systems is the key to understanding many of their properties. In this paper we examine the free energy landscape of the Ising Sherrington-Kirkpatrick (SK) model \cite{sherrington:75} within the framework of the Thouless-Anderson-Palmer (TAP) equations \cite{thouless:77}. The free energy landscape according to the TAP equations is strikingly simple \cite{aspelmeier:04,cavagna:04}. For each solution of the TAP equations, which is a minimum of the free energy, there is an associated saddle point (which has one negative eigenvalue in its Hessian), and the number of these pairs of stationary points is exponentially large in $N$, the number of spins in the system \cite{Bray_1980, Bray:81, Bray:81b}. The barrier to escape from the minimum is just the difference in free energy between the saddle point and the minimum. In Ref. \cite{aspelmeier:04} it was shown how one could numerically obtain the saddle point starting from the minimum. Solutions of the TAP equations with free energies per spin $f$ lying within $O(1/N)$ of the free energy $f_0$ of the state of lowest free energy correspond to pure states \cite{bray:84}. A feature of the TAP equations is the existence of a critical free energy $f_c$ above which the solutions have zero overlap with each other \cite{Bray:81b,Bray:81,bray:84}, whereas in the interval $f_0 \le f < f_c$, the TAP solutions  have overlaps with each other similar to those in the Parisi replica symmetry breaking (RSB) solution \cite{parisi:79,parisi:83,rammal:86,mezard:87,parisi:08}.

Right from the earliest days of finding numerical solutions of the TAP equations it has been observed that the solutions which are found have \textit{marginal} stability \cite{Bray:78a}. Marginal stability is also found in a wide range of physical systems \cite{muller:15}. We shall define marginal to mean that their Hessian eigenvalues, calculated at the TAP minimum, have a distribution which has support all the way down to zero. This is surprising since for the overwhelming majority of TAP minima (that is, those whose $f>f_c$) there is a gap in their Hessian spectrum \cite{Bray_1980,Bray:81,Bray:81b,aspelmeier:19}. Why is it then that numerical work for large values of $N$ does not find these states, (although they can be found for small values of $N$ \cite{cavagna:04})? We believe that our work in this paper provides the explanation of this long-standing puzzle. We shall also explain why numerical work can find states with  $f < f_c$ (and in fact quite close to to $f_0$,  the free energy of the pure states) \cite{aspelmeier:06,aspelmeier:19},   which seem not to have  RSB features. There is again a paradox:  States at free energies $f < f_c$ without RSB features must be exponentially rare in comparison to states which would have RSB features of their overlaps. Nevertheless we shall give an argument  in  Sec. \ref{RSstates} that such states must exist. These states are the ones which are found via numerical solutions of the TAP equations. The \lq\lq Edwards" procedure \cite{baule:18} of determining the complexity of the TAP states at a given free energy $f$ leads to the prediction of a critical free energy $f_c$ \cite{Bray:81b, Bray:81}, but does not describe well what one sees in numerical solutions of the TAP equations. In fact, what one sees is very similar to the behavior found in quenches from infinite temperature to low temperatures \cite{aspelmeier:19}. Nevertheless the Edwards procedure seems to be the only way one can do analytical calculations and we shall use it extensively in this paper. The Edwards procedure for calculating a quantity gives its average over \textit{all} the TAP states at a specified free energy $f$, averaged over the spin couplings $J_{ij}$.

The height of the barrier between the minimum and the saddle point has a probability distribution. A full treatment would involve calculating the form  of this distribution. In this paper we have a more modest goal, which is to establish the $N$ dependence of the typical barriers at particular values of $f$, when averaged over $J_{ij}$.  We find the answer depends on the regime, $f_0 \le f< f_c$, $f=f_c$  or $f> f_c$. For $f > f_c$ the barriers at large $N$ are very small as they decrease as $1/N^2$. Right at $f = f_c$ the barriers are of $O(1)$.  The barriers of TAP states with $f >f_c$ are so small  that these states will have no dynamical significance. In fact in this region the barriers are such that when solving the TAP equations at large values of $N$  the iterations typically take one step towards but beyond the minimum and right over its accompanying saddle towards the trivial and unphysical  minimum at all $m_i =0$. It is this fact which  explains why TAP solutions with large $N$ at $f > f_c$ are just not found. We are confident of these results as they can be supported via direct calculations of finite size corrections \cite{Owen:82} and direct solution of the TAP equations \cite{aspelmeier:19}.

For $f < f_c$ our results are only tentative. A little progress has been made using a mixture of old arguments \cite{Bray:81b,Bray:81,dasgupta:83} together with scaling arguments \cite{aspelmeier:08}. One possibility  that emerges is that for all $f \le f_c$ the typical barriers are of order $N^{1/3}$. One of the
key questions in the theory of spin glasses is the $N$ dependence of the  barriers separating pure states.  The picture of an ordered state consisting of many pure states   comes from the  Parisi \cite{parisi:79,parisi:83,rammal:86,mezard:87,parisi:08} replica symmetry breaking (RSB) picture of spin glasses. This is a picture which has been established for mean-field calculations of the equilibrium state, and is valid for the SK model.
In the RSB picture the many pure states present have free energies which differ by $O(1)$. Unless the barriers between them become infinite in the thermodynamic limit of $N\to \infty$, the pure states will not be well-defined: If they are finite, thermal fluctuations would mix the pure states together and the RSB picture of many pure states would not be possible.
There are old arguments \cite{Rodgers:89,aspelmeier:06,Kinzelbach:91} suggesting that in the SK model these barriers could depend on the number of spins $N$ as $N^{1/3}$, which are at least consistent with the results of simulations \cite{Billoire:10, billoire:01,Bittner:06,Monthus:09,Colborne:90}.  Alas neither  the arguments nor the simulations can at the present time be regarded as conclusive. In fact the most recent and extensive simulations \cite{Bernaschi:20} suggest that the exponent may even be smaller than $1/3$.

The extensive simulations of  Bernaschi et al. \cite{Bernaschi:20} were not only done for  the SK model but also for another mean-field model, the Viana-Bray model \cite{viana:85}. For the latter they could take advantage of the fact that each spin in only coupled to a finite number of other spins to study systems with very large values of $N$. They found that their results seemed to be at variance with the expectations of Kurchan and Cugliandolo \cite{cugliandolo:93,cugliandolo:94} who argued that at least for the SK model the dynamics at long times would not be determined by the initial conditions.   Bernaschi et al. \cite{Bernaschi:20} found instead that for the Viana-Bray model the system remained trapped in the vicinity of its initial state for temperatures below the transition temperature: The system was therefore non-ergodic. They were unsure whether their results would extend to the SK model for there, because each spin is coupled to all the other $N-1$ spins the computation is slow which prevented them studying  large values of $N$. It would be rather disconcerting if two different types of mean-field model were to give fundamentally different results. We wanted to determine the barriers for $f < f_c$ for the SK model to see if there were large barriers for all TAP states with such free energies, just as there must be between the pure states. Our conclusion in this study is that indeed the barriers might be large for $f < f_c$ (of order $N^{1/3}$) when the SK model would also be non-ergodic just like the Viana-Bray model. In Sec. \ref{barrierslessfc} other possibilities for $f < f_c$ are also outlined.

In this paper the critical value $f_c$  plays a prominent role. TAP solutions with free energies $f > f_c$ have no overlap with each other, while those with free energies $f < f_c$ have overlaps with each other and when constructing the Edwards complexity average one needs replica symmetry breaking techniques when $f < f_c$ \cite{Bray:81, bray:84, muller:06}. However, in numerical studies $f_c$ seems to be invisible \cite{aspelmeier:06, aspelmeier:19}: One just converges to TAP solutions at free energies $f < f_c$ which have no overlap with each other. The actual value which they converge to depends on the numerical technique used \cite{aspelmeier:06, aspelmeier:19}, but is always lower than $f_c$. In fact, in Ref. \cite{aspelmeier:06} we found methods which yielded TAP solutions which were very close to $f_0$. Why such solutions with no replica symmetry breaking of their overlaps  can exist will be explained in Sec. \ref{RSstates}.

In Sec. \ref{TAP} we  present a formalism  for the calculation of the barrier height for a TAP solution.  To do this we shall focus on the Taylor series expansion in $q$ about a TAP (minimum) solution (see Eq. (\ref{Taylor})), and work out the coefficient (called $c$) of the cubic term  and the quadratic term (called $a$) in the expansion. In the large $N$ limit, this is sufficient to determine the (barrier) height of the saddle point above the minimum, provided the $N$ dependence of $a$ and $c$ can also be determined. Other procedures have been used for studying barriers etc. in spin glasses, notably the comprehensive paper of  Ref. \cite{muller:06}, which involved the use of both replicas and two-group replica symmetry breaking \cite{Bray:1979,potters:95}. Supersymmetry methods have also been used \cite{parisi:04,rizzo:05}.  However, all these methods run into similar difficulties when finding the $N$ dependence of the barrier heights i.e. the $N$ dependence of the coefficients $a$ and $c$.

In Sec. \ref{results} we present the main results of our calculations. The actual calculations are tedious and lengthy so we have relegated them to four Appendices. In Appendix \ref{sec:complexity} we describe once more the Edwards style calculation of  the complexity of the TAP solutions. In Appendix \ref{2beta2H} we show that in the thermodynamic limit the coefficient $a$ is $0$  (but when $f > f_c$ it is of magnitude $1/N$, as demonstrated in Sec. \ref{results}). Appendix \ref{norm} provides a result needed in the calculation of the cubic term $c$, which is done in Appendix \ref{ccalc}.

\section{The TAP equations and free energy}
\label{TAP}
In this section we present the TAP equations and define some of the quantities needed to calculate  the barriers such as the coefficients $a$ and $c$.
\begin{figure}
  \includegraphics[width=\columnwidth]{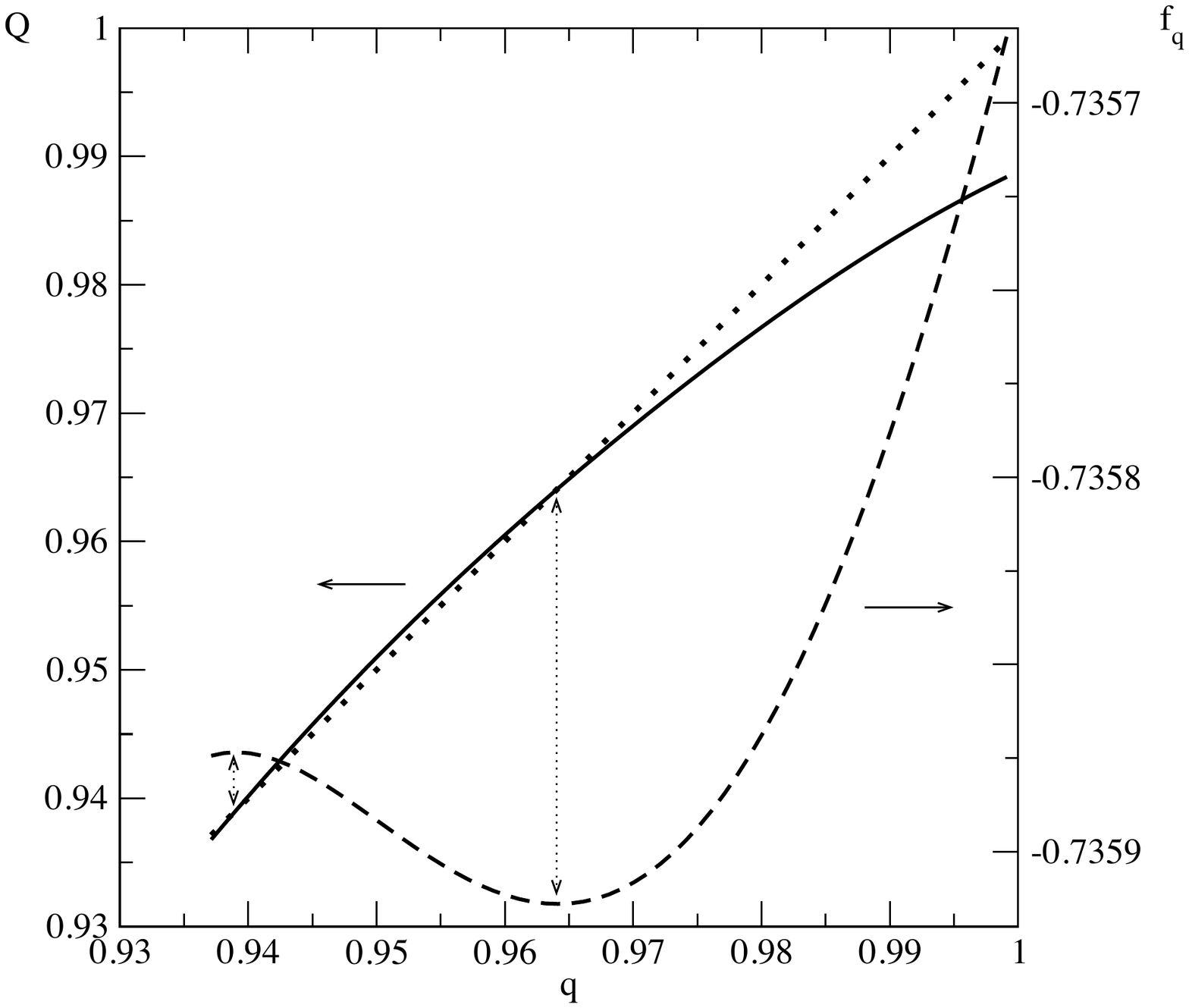}
  \caption{The functions $Q(q)$ (continuous line) and the free energy per spin, $f_q=F(q)/N$, (dashed line) associated with a particular TAP solution. The minimum and the saddle point occur where $Q(q)$ crosses the dotted line $Q = q$. The figure was obtained for $N = 200$ spins at a temperature $T= 0.2T_c$, and $T_c=1$. At values of $q$ somewhat smaller than that at the saddle of $f_q$ (which appears as a maximum here), the TAP equations lose their validity \cite{Owen:82,thouless:77}.}
  \label{fig:Q(q)}
\end{figure}
Our treatment follows closely the procedure which we used in Ref. \cite{aspelmeier:04}.
We write the TAP free energy (multiplied by $\beta= 1/(k_B T)$) as
\begin{eqnarray}
F_q(m_i) &=& -\frac{\beta}{2}\sum_{i,j}J_{ij}m_im_j-\frac{N}{4}\beta^2(1-q)^2
- N \ln 2  \nonumber \\
 &+& \sum_i\left(\frac{1}{2}\ln(1-m_i^2) + m_i\tanh^{-1}m_i\right)
 \nonumber\\&+&  \frac{1}{2} \beta^2(1-q) \big(\sum_i m_i^2-Nq\big).
\label{FTAP}
\end{eqnarray}
The first two lines are just the conventional form of the TAP free energy \cite{thouless:77}, if $q$ is defined to be $q=\sum_i m_i^2/N$. Instead the functional of Eq. (\ref{FTAP}) consists of $(N+1)$ variables, the $m_i$ \textit{and} $q$. Stationarity with respect to $m_i$ gives the TAP equations
\begin{eqnarray}
&&\partial F_q(m_i)/\partial m_i\equiv G_i \nonumber\\&=& -\beta \sum_j J_{ij}m_j+\tanh^{-1}m_i +\beta^2(1-q)m_i=0.
\label{tap}
\end{eqnarray}
We then take the solutions of these $N$ stationarity equations, $m_i(q)$ and construct the functions $F(q) \equiv F_q(\{m_i(q)\})$ and $Q(q)=\sum _i m_i(q)^2/N$. One readily verifies that the stationarity equation for $F(q)$ reproduces the standard TAP equations, which are given by Eq. (\ref{tap}) but with $q$ \textit{defined} to be $\sum_im_i^2/N$. An example of $F(q)$ for a particular bond realization and $N= 200$ at temperature $T = 0.2T_c$ is plotted in Fig.\ref{fig:Q(q)}. It is a concrete realization of the schematic figure in the original TAP paper \cite{thouless:77}. The additional stationarity equation of Eq. (\ref{FTAP}),
\begin{equation}
0=\frac{\partial F_q(m_i)}{\partial q}=(\beta^2/2)(Nq-\sum_i m_i^2)
\label{firstdiffq}
\end{equation}
forces $Q = q$ at the stationary points in the full $(N+1)$ dimensional space. Thus at the minimum and saddle-point of the  free energy function of Eq. (\ref{FTAP})  coincides with  that for the free energy of the original TAP free energy. The free energy barrier
is then just the difference in free energies between the saddle-point and the minimum.

It is useful to introduce the matrix
\begin{equation}
(X^{-1})_{ij}= \frac{\partial G_i}{\partial m_j}=\big[\frac{1}{1-m_i^2}+\beta^2(1-q)\big]\delta_{ij}-\beta J_{ij},
\label{Xdef}
\end{equation}
and the function $g(m_i)$ by
\begin{equation}
g(m_i)= \tanh^{-1}m_i +\beta^2 (1-q) m_i.
\label{gdef}
\end{equation}
The susceptibility matrix of the original TAP equations, (that is when $q$ is defined to equal $\sum_i m_i^2/N$) is $(A^{-1})_{ij}=\partial m_i/\partial h_j$. It gives the response of the $m_i$ to an infinitesimal site dependent field $h_j$. It can be written as a sum of $O(1)$ terms, involving $X_{ij}$ plus a term of order $1/N$:
\begin{equation}
A_{ij}= (X^{-1})_{ij}-\frac{2 \beta^2}{N} m_i m_j.
\label{Adef}
\end{equation}
The term of order $1/N$ plays a very important role \cite{aspelmeier:04}.

We shall now obtain expressions for the first three derivatives of $F(q)$ at its minimum. Thus we are expanding about the minimum in Fig. \ref{fig:Q(q)}. The saddle point is the maximum of the function $F(q)$ in that figure. We have shown that when $N$ is large it is sufficient just to determine the first three derivatives to calculate the barrier height \cite{aspelmeier:19}. In Eq. (\ref{FTAP}) we were regarding $\{m_i\}$ and
$q$ as independent variables. However, when expanding the free energy about its the minimum, the $\{m_i\}$ at the minimum are  $q$  dependent, because  the TAP equations $\{G_i=0\}$ link them.

{\it The first derivative}

The function $F(q)$ has first derivative
\begin{eqnarray}
dF(q)/dq&=&\sum_i \frac{\partial F_q(\{m_i\})}{\partial m_i}\,\, \frac{\partial m_i}{\partial q}+\frac{\partial F_q(\{m_i\})}{\partial q} \nonumber \\
&=&\sum_i G_i \frac{\partial m_i}{ \partial q}+\frac{\partial F_q(\{m_i\})}{\partial q}.
\label{1Pderiv}
\end{eqnarray}
From Eq. (\ref{FTAP}) we have
\begin {equation}
\frac{\partial F_q(\{m_i\})} {\partial q}=\frac{1}{2} \beta^2 (N q -\sum_i m_i^2).
\label{0deriv}
\end{equation}
On using Eq. (\ref{tap}) one can see that the first term in Eq. (\ref{1Pderiv}) is zero.
Hence
\begin{equation}
\frac{dF(q)}{dq}=\frac{\beta^2}{2}  (N q -\sum_i m_i^2).
\label{1deriv}
\end{equation}
Thus at stationary points where $Q=q$, this derivative is  zero according to Eq. (\ref{firstdiffq}).

{\it  The second derivative}

Differentiating Eq. (\ref{1Pderiv}) with respect to $q$
\begin{eqnarray}
\frac{d^2 F(q)}{d q^2}&=& \sum_{i}\big( \frac{d G_i}{dq} \,\,\partial m_i/\partial q
+\sum_i G_i \partial^2 m_i/\partial q^2\big) \nonumber \\&+& d  (\frac{\partial F_q(\{m_i\})}{\partial q})/d q .
\label{2Pderiv}
\end{eqnarray}
Because $G_i(q) = 0$ for all values of $q$, total derivatives like $d^n G_i/d q^n=0$, for any value of $n$. Thus the top line of Eq. (\ref{2Pderiv}) gives zero.

From Eq. (\ref{1deriv})  the second line of Eq. (\ref{2Pderiv}) is
\begin{equation}
d (\frac{\partial F_q(\{m_i\})}{\partial q})/d q=\frac{1}{2} \beta^2(N-2 \sum_i m_i \partial m_i/\partial q).
\label{2partial}
\end{equation}
Note that the partial derivative of $G_i$ with respect to $q$ is
\begin{equation}
\partial G_i/\partial q=-\beta^2 m_i.
\label{gderiv}
\end{equation}
Hence
\begin{equation}
 \partial m_i/\partial q \equiv v_i = \beta^2 \sum_j X_{ij} m_j,
 \label{vdef}
 \end{equation}
 which follows from differentiating the stationarity  equation $G_i=0$ with respect to $q$ and noting that $dG_i/dq=0$ at the stationary point.
 Thus the second derivative is finally
 \begin{equation}
 \frac{d^2 F(q)}{d q^2}=\frac{1}{2} \beta^2(N-2 \sum_i m_i \partial m_i/\partial q),
 \label{2deriv}
 \end{equation}
 which is equivalent to
 \begin{eqnarray}
 d^2 F(q)/d q^2 &=&\frac{N \beta^2}{2} (1-\partial Q/\partial q) \nonumber \\
 &=& \frac{N \beta^2}{2}(1-2 \beta^2 H)\equiv N a,
 \label{Xform2deriv}
 \end{eqnarray}
 where
 \begin{equation}
 H=\frac{1}{N}\sum_{i,j} m_i X_{ij} m_j.
 \label{Hdef}
 \end{equation}
 In Appendix \ref{2beta2H}  we will show that $(1-2 \beta^2 H)$ vanishes as $N \to \infty$. For finite values of $N$ it is of order $1/N$ when $f >f_c$. We shall argue that it is of order $1/N^{1/3}$ for all $f \le f_c$.

 {\it The third derivative}

Using Eqs. (\ref{2Pderiv}) and (\ref{2partial}) the third derivative can be seen to be
\begin{multline}
d^3 F(q)/dq^3=-\beta^2 \sum_i \partial (m_i \partial m_i/\partial q)/\partial q  \nonumber\\
= -\beta^2 \sum_i (\partial m_i/\partial q)^2-\beta^2 \sum_i m_i \partial^2 m_i/\partial q^2.
\end{multline}
From Eq. (\ref{vdef}),
\begin{equation}
\sum_j X^{-1}_{ij} \partial m_j/\partial q=\beta^2 m_i
\end{equation}
so
\begin{multline}
\beta^2 \partial m_i/\partial q=\sum_j\bigg[ \delta_{ij}\bigg ( \frac{2 m_i} {(1-m_i^2)^2}\partial m_i /\partial q -\beta^2\bigg) \partial m_j/\partial q\nonumber\\ +X^{-1}_{ij} \partial ^2 m_j/\partial q^2\bigg].
\end{multline}
Hence
\begin{equation}
\partial^2 m_i/\partial q^2=  \sum_{j} X_{ij} \bigg(2 \beta^2 \partial m_j/\partial q- \frac{2m_j}{(1-m_j^2)^2}(\partial m_j/\partial q)^2\bigg).
\end{equation}
Hence
\begin{multline}
d^3 F(q)/dq^3=N\bigg(-\frac{3 \beta^2}{N}\sum_i v_i^2 + \frac{1}{N} \sum_i \frac{2 m_i v_i^3}{(1-m_i^2)^2}\bigg)\\\equiv N c.
\label{3deriv}
\end{multline}
We shall show in Appendix \ref{ccalc} that $c$ is of $O(1)$ for $f \ge f_c$.

To summarise: The TAP free energy landscape for the Ising SK spin glass is very simple. It consists of an exponentially large number of minima and their associated index one saddles. The barrier height is the difference in free energy between the saddle-point and the minimum, as in Fig. \ref{fig:Q(q)}. For $N$ large, it is possible to obtain this height from the second and third derivatives of the free energy, $a$ and $c$, calculated at the minimum $q=q_m$, if $N$ is large,
\begin{equation}
F(q)-F(q_m)= N\bigg(\frac{a}{2}(q-q_m)^2+\frac{c}{6}(q-q_m)^3\bigg).
\label{Taylor}
\end{equation}
The coefficient
$a$ is given in Eq. (\ref{Xform2deriv}) and the coefficient $c$ is given by Eq. (\ref{3deriv}).  One can determine the $N$ dependence of the barrier heights $B$ if one knows the $N$ dependence of the coefficients $a$ and $c$ and this is what is discussed in the next section.

\section{Barrier heights for $f\ge f_c$}
\label{results}
In this section we state our main results for the region $f \ge f_c$. Our basic approach is to calculate the coefficients $a$ and $c$ using the methods previously employed to obtain the complexity (the calculation of which is briefly described in Appendix \ref{sec:complexity}).

The value of $q$ at the saddle point, $q_s$, can be calculated by finding when $dF(q)/dq=0$ in Eq. (\ref{Taylor}), and is
\begin{equation}
q_s =q_m-\frac{2 a}{c}.
\label{qsdef}
\end{equation}
Notice that because $a$ and $c$ are positive $q_s < q_m$ and this feature is also visible in Fig.\ref{fig:Q(q)}.

 We shall find that at least for minima whose $f \ge f_c$, that the cubic coefficient $c$ is finite and of $O(1)$ and right at $f =f_c$ takes the value $2.439723 \beta^2$ according to Appendix \ref{ccalc}.

    The quadratic coefficient $a$ is of order $O(1/N)$ for $f> f_c$ (see Eq. (\ref{fabovefc})) and is $O(1/N^{1/3})$ at $f = f_c$  (see Eq. (\ref{fatfc})).  Then
$q_s$ is less than  $q_m$ by $O(1/N)$ for $f > f_c$.
Thus in the large $N$ limit the saddle and the minimum will merge together.

The higher derivatives of $F(q)$ have been neglected in Eq. (\ref{Taylor}). The extent to which this is a good approximation for values of $N < 320$ is discussed to some extent in  \cite{aspelmeier:19}. For large values of $N$, when $q_s \to q_m$, it could be expected to be an excellent approximation. The barrier height $B$ is then
\begin{equation}
B \equiv F(q_s)-F(q_m)= N \frac{2 a^3}{3 c^2}.
\label{Bdef}
\end{equation}
Hence $B$ is of $O(1/N^2)$ for $f>f_c$ but right at $f_c$ it is of $O(1)$.  Our numerical work suggest that once over the saddle in the direction away from the minimum one often plunges down towards the paramagnetic solution of the TAP equations,  $q=0$ and $m_i=0$.  This has a lower free energy that the minimum when $T < T_c$ but lies however in the region of parameter space where the TAP equations have no validity \cite{thouless:77, Owen:82, Plefka:02}.

In Ref. \cite{aspelmeier:04}, (see also \cite{parisi:04}) we suggested that $v_i \equiv \beta^2 \sum_j X_{ij} m_j$ was proportional  the lowest eigenvalue of the Hessian matrix (inverse susceptibility matrix)  $A_{ij}$,
where $A_{ij}=(X^{-1})_{ij}-2 \beta^2 m_i m_j/N$. The smallest eigenvalue of $\mathbf{A}$, $\lambda_{min}$, must be such that
\begin{equation}
\lambda_{min} \le \frac{\sum_{ij} v_i A_{ij} v_j}{\sum_i v_i^2}.
\label{bound}
\end{equation}
Then
\begin{equation}
\lambda_{min} \le\frac{\beta^2 H(1-2 \beta^2 H)}{\sum_i v_i^2/N}.
\label{lambdamin}
\end{equation}
Because the coefficient $a$ of the quadratic term in Eq. (\ref{Taylor}) is also proportional to $(1-2 \beta^2 H)$, Eq. (\ref{lambdamin}) indicates that the route from the minimum to the saddle point must be starting from the minimum in the direction of the smallest eigenvector $v_i$.
For TAP solutions with $f >f_c$ we argue below  that the finite size scaling form is
\begin{equation}
(1-2 \beta^2 H) \sim \frac{1}{N (f-f_c)^2},
\label{fabovefc}
\end{equation}
when $N(f-f_c)^3>>1$. In the opposite limit $N(f-f_c)^3\to 0$,
\begin{equation}
(1-2 \beta^2 H) \sim \frac{1}{N^{1/3}}.
\label{fatfc}
\end{equation}
In Appendix \ref{norm}  we show that $\sum_i v_i^2/N \sim 1/x_p$, where
\begin{equation}
x_p=1-\frac{\beta^2}{N}\sum_i (1-m_i^2)^2.
\label{xpdef}
\end{equation}
As $f$ approaches $f_c$, $x_p \propto (f-f_c)$ and right at $f_c$, $x_p \sim 1/N^{1/3}$. ($f_c$ is determined by finding where $x_p$ becomes zero as $f$ is decreased).  The finite size scaling form is
\begin{equation}
x_p=(f-f_c) \mathcal F((f-f_c)N^{1/3}),
\label{xpscaling}
\end{equation}
for $f >f_c$: The crossover function $\mathcal F(x)$ goes to a constant as $x\to \infty$ and goes like $1/x$ as $x\to 0$ so right at $f = f_c$, $x_p \sim 1/N^{1/3}$.
We suspect that it has the same $1/N^{1/3}$ dependence too for all $f < f_c$, right down to and including the pure states.

Then using Eq. (\ref{lambdamin}) the smallest eigenvalue of the Hessian matrix for $f> f_c$ is
\begin{equation}
 \lambda_{min} \sim \frac{1}{N (f-f_c)},
 \label{minflarge}
 \end{equation}
  which is a "null" eigenvalue in the large $N$ limit. The other $N-1$ eigenvalues are separated from it by a finite gap $x_p^2/(4p)$ \cite{Bray:78a,Plefka:02,aspelmeier:19}, where
\begin{equation}
 p=\frac{\beta^3}{N}\sum_i (1-m_i^2)^3.
\label{pdef}
\end{equation}
$p$ is finite at $f=f_c$.
The null eigenvalue is a consequence of a broken supersymmetry \cite{parisi:04}. The scaling form of $\lambda_{min}$ as $f \to f_c$ is
\begin{equation}
\lambda_{min}= \frac{1}{N(f-f_c)}\tilde{\mathcal F}((f-f_c)N^{1/3}),
\label{lambdamincrossover}
\end{equation}
which gives $\lambda_{min}\sim 1/N^{2/3}$ right at $f=f_c$.
For $f > f_c$,  there is a finite band gap above the null eigenvalue starting at $x_p^2/(4p)$ \cite{aspelmeier:19} which using the crossover form is of order $1/N^{2/3}$ right at $f=f_c$. Thus for $f= f_c$ the band gap disappears and the null eigenvalue  becomes just the lowest eigenvalue of the band.

These estimates are consistent with the density of states of the $\mathbf{A}$ matrix, which  is of the form $\rho(\lambda) \approx D \sqrt{\lambda}$ for small $\lambda$ \cite{Bray:78a} at $f =f_c$.  One can obtain $\lambda_{min}$ via
 \begin{equation}
 1 = N\int_0^{\lambda_{min}} d\lambda \,D \sqrt{\lambda}
 \label{sqrtlambda}
 \end{equation}
 which also gives $\lambda_{min} \sim 1/N^{2/3}$.  This is consistent with the band-edge estimate $x_p^2/(4p)$ as $x_p \sim 1/N^{1/3}$ at $f=f_c$.  We would also expect the same form for $f < f_c$ as at $f = f_c$,  that is  $\lambda_{min} \sim 1/N^{2/3}$ and that the coefficient $a$ of Eq. (\ref{Taylor}) is also $ a \sim 1/N^{1/3}$.

 We next explain why the coefficient $a$ is of order $1/N$ for $f > f_c$. To obtain this result we have to use the leading correction to the TAP free energy \cite{Owen:82, thouless:77}. (Note that this is "controversial"; Plefka \cite{Plefka:02,Plefka:20} has long advocated  different corrections which we have discussed before \cite{aspelmeier:04, aspelmeier:19}).  Owen's correction \cite{Owen:82} is,
\begin{equation}
F= F_q(m_i)-\frac{1}{4} \ln[x_p].
\label{corr}
\end{equation}
The term $F_q(m_i)$ is $O(N)$ while the correction term is of $O(1)$. There are other correction terms which are negligible in the finite size scaling limit $N x_p^3\sim N (f-f_c)^3$ of $O(1)$ as $N >> 1$. (This can be compared to the finite size scaling combination near $T_c$ of the SK model $N \tau^3$ where $\tau= T/T_c-1$ \cite{aspelmeier:08}). The $1/N$ correction to the coefficient $a$ is then
\begin{equation}
a =\frac{1}{4N}\bigg(\frac{1}{x_p^2}\big(\frac{\partial x_p}{\partial q}\big)^2-\frac{1}{x_p} \frac{\partial^2 x_p}{\partial q^2}\bigg).
\end{equation}
The derivatives $\partial x_p/\partial q$ and $\partial^2 x_p/\partial q^2$ are finite as $x_p\to 0$. For example, $\partial x_p/\partial q\to 2-6 q$. Hence the finite size scaling form of $a$ is as $\sim 1/(N x_p^2)$ for $N x_p^3>>1$. This leads to the barriers $B$ being as small as  $O(1/N^2)$ for $f > f_c$. It is the existence of such small barriers when $f > f_c$ at large values of $N$ which prevents one finding numerical solutions of the TAP equations in this free energy range.

To summarise: for TAP states with $f> f_c$ the barriers are of $O(1/N^2)$, and are of $O(1)$ at $f= f_c$.

\section{Barriers for $f < f_c$}
\label{barrierslessfc}
 For $ f < f_c$ (that is, for solutions which have RSB like overlaps  with other solutions of free energy $f$) there exists no information from direct solutions of the TAP equations to guide us. However, it would be possible (at least in principle) to extend the calculations presented in Appendices A, B, C, and D into this regime. In fact the equations just for the calculation of the complexity (the analogues of those in Appendix A) were written down long ago \cite{Bray:81, bray:84}. Solving these equations is very difficult and has never been achieved \cite{muller:06}, and the only success has been for the limit $f \to f_0$. Then the solution has very similar features to the Parisi form of $q(x)$, $x \in[0,1]$ \cite{mezard:87}. In the opposite limit of $f \to f_c$, we suspect that the solutions go over to those with replica symmetry with the "breakpoint" $x_1$ going to zero in this limit.

 However, Eq. (\ref{qsdef}) does lead to some information. We  have suggested that for $f < f_c$ that $a$ is of order $1/N^{1/3}$, so we need to find what happens to $c$. When $f\ge f_c$ we  show in Appendix \ref{ccalc} that $c$ remains finite.
A possibility for $f < f_c$ is that  $c =  h(f_c-f)N^{1/6})$, where the crossover function $h(x)$ is of $O(1)$ for $x\to 0$, in order to go to a constant at $f=f_c$, but decreases as $1/x$ at large $x$. (The scaling combination of $(f_c-f)$ and $N$ used here parallels that for the number of steps of replica symmetry breaking \cite{aspelmeier:08}, where we have changed $(T_c-T)$ to $(f_c-f)$ as was suggested in \cite{Bray:81b}). This would make $c$ of $O(1/N^{1/6})$ as $N \to \infty$ at fixed any fixed $f < f_c$.

Support for this possibility comes from an old result of Dasgupta and Sompolinsky \cite{dasgupta:83}.  One can write
\begin{equation}
Q = \frac{1}{N} \sum_i m_i^2=\frac{1}{N}\sum_{\lambda} m_{\lambda}^2,
\end{equation}
where we have expressed the magnetizations in terms of the eigenvectors of the matrix $A_{ij}$. (Actually the authors of Ref. \cite{dasgupta:83} used instead the eigenvectors of $J_{ij}$, but $A_{ij}$ is the better choice for our argument). We are arguing that in the change from the minimum to the saddle goes along the lowest eigenvector of the $A_{ij}$. The Dasgupta and Somplinsky argument, which relies on the use of replica symmetry breaking, suggests that this lowest eigenvector made a contribution to $Q$ of order $m_{\lambda_{min}}^2/N\sim 1/N^{1/6}$, which means that $c$ must be of magnitude $O(1/N^{1/6})$, on using Eq. (\ref{qsdef}), when $a$ is of $O(1/N^{1/3})$. Then $q_m-q_s  \sim 1/N^{1/6}$.

In the replica treatment of the complexity there are functions
$\eta(x)$ and $\eta^*(x)$ \cite{Bray:81, bray:84}.  $\eta(1)=\beta^2/N\sum \langle m_{\lambda}^2\rangle$, where here $\langle \cdots\rangle $ denotes an average over all TAP solutions \cite{Bray:81}. In the opposite limit of $x \to 0$, $\eta(0)$ is the average over the most "distant" solutions in solution space and it is those which Dasgupta and Sompolinsky focussed upon.

If $c$ does go at large $N$ as $1/N^{1/6}$ for $f< f_c$  then the TAP states with $f < f_c$ will have barriers which scale as $N^{1/3}$, as can be seen using Eq. (\ref{Bdef}) if $a\sim 1/N^{1/3}$.

With so many states having large barriers (there are an exponentially large number of states with $f < f_c$), the SK model would have the the same non-ergodic properties as Bernaschi et al.         \cite{Bernaschi:20} found in the Viana-Bray model. The dependence of $c$ on $N$ as $1/N^{1/6}$ indicates that it is zero in the thermodynamic limit and suggests that by generalizing the calculations of Appendix D for $f < f_c$ and including the consequences of replica symmetry breaking it might be possible to actually prove it. This is worth considering in light of the importance of explaining the simulation results of  Ref. \cite{Bernaschi:20}, but would be very challenging \cite{muller:06}.

Another possibility could be that $c$ stays finite when averaged over all states of free energy $f$ and only goes to zero for the pure states. If $c \sim (f-f_0)^{1/6}$, then pure states which have free energies per spin $f$ larger than $f_0$ by an amount of $O(1/N)$, would be associated with barriers which grow as $N^{1/3}$ \cite{aspelmeier:06}. If this is the correct possibility, then the SK model would behave as suggested by Cugliandolo and Kurchan \cite{cugliandolo:93,cugliandolo:94}. Another possibility could be that $c$ stays finite on average but it acquires a very wide distribution so that at a subset of the TAP minima the coefficient $c$ is very small, of order $1/N^{1/6}$. To investigate which, if any, of these possibilities is correct requires the extension of the calculations in Appendix \ref{ccalc} into the region where $f < f_c$, and the use of replica symmetry breaking and the incorporation of finite $N$ effects.

The coefficient $a$ always is small, dependent on some inverse power of $N$ in the large $N$ limit. Its small value is related to the existence of the null eigenvalue. The existence of the null eigenvalue is in a sense obvious \cite{parisi:04, rizzo:05}. This is because the complexity, which is the log of the number of TAP solutions is a function of temperature $T$ and free energy per spin $f$. Any increase  in, say $T$ will cause an exponentially large decrease in the number of TAP solutions. A TAP solution, defined here as the minimum and its associated saddle point, will disappear through the merging of the saddle and the minimum, which happens if  $a=0$. Thus the coefficient $a$ must for any given solution have a value taken from its probability distribution, and this value gives us via Eq. (\ref{qsdef}) the value of $q_s-q_m$. In Appendix \ref{ccalc} we obtain its average value for $f \ge f_c$ by using the Edwards average over the bonds $J_{ij}$ of all solutions of free energy $f$. The much harder task of studying the distribution of $c$ and the distribution of  the barriers when $f < f_c$ is a challenge for the future.

\section{TAP solutions without RSB overlaps for $f < f_c$}
\label{RSstates}
Our studies in Refs. \cite{aspelmeier:06} and \cite{aspelmeier:19} showed that numerical methods existed which produced solutions with free energies per spin $f<f_c$ (some gave results close to $f_0$),  but no signs of replica symmetry breaking of their overlaps.   One might wonder whether replica symmetric states with $f< f_c$ should even exist, given that there are an exponentially large number of solutions with $f < f_c$ with RSB features. We next give an argument that replica symmetric states must exist with free energies at free energies $f < f_c$.

Our argument starts with an old paper of Bray \cite{Bray_1982}. Bray asked what is the ordering field of the spin glass. When the ordering field is applied to a system undergoing a transition there is no phase transition, as for example in a ferromagnet in a uniform field. In a spin glass a uniform field only suppresses the transition down to the de Almeida-Thouless line \cite{dealmeida:78}. However Bray discovered that application of a field along the largest eigenvector of the $J_{ij}$ matrix suppressed the transition to much lower temperatures than the application of a uniform field. The suggestion we would make is that a field along the lowest eigenvector of the $A_{ij}$ matrix is the ordering field. This is also the eigenvector (see Sec. \ref{results})
along which one passes from the minimum to the saddle and is the "null" eigenvector associated with the broken supersymmetry \cite{parisi:04,rizzo:05}.

In the presence of the ordering field one is always in the paramagnetic phase. In the paramagnetic phase the complexity of the TAP solutions is zero. In zero field there is but a unique solution, all $m_i=0$. In the presence of the ordering field there might be multiple solutions even if they are not exponentially numerous (but we have only ever found just one). Thus by reducing the temperature towards zero in the presence of the ordering field and then by turning off the ordering field one should be able to reach the ground-state of the SK model (for any realization of the bonds!). (Since this is an NP hard problem, something must happen to make this impossible, but we have not discovered what it might be).  However, the argument does illustrate that families of states with replica symmetry must exist for $f < f_c$.

\acknowledgments
We should like to thank Prof. F. Guerra for a discussion of finite size effects.

\appendix

\section {Calculation of the Edwards average of the complexity}
\label{sec:complexity}

The formulae of Eq. (\ref{Xform2deriv}) for the coefficient $a$ and Eq. (\ref{3deriv}) for the coefficient $c$ refer to a single solution of the TAP equations. It will explicitly depend on the bonds $J_{ij}$, and the only way one can make progress analytically is by averaging over the bonds. These calculations then become variants of those used long ago for the complexity \cite{Bray_1980, Bray:81, Bray:81b}, which is related to $N_s(f)$, the number of solutions (per unit free-energy range) with free energy $f=F/N$, scaled by $\beta$). These calculations will just  be briefly summarized in this Appendix.  $N_s(f)$ is given by
\begin{eqnarray}
N_s(f)& \equiv& \int W= N^2\int_0^1 dq \int_{-1}^{1}(d m_i) \delta \bigg(Nq-\sum_i m_i^2\bigg) \nonumber \\&\times & \delta \bigg(N f -\sum_i f_1(m_i,q)\bigg)\prod_i \delta (G_i)|\mathrm{det}\mathbf{A}|,
\label{Nsdef}
\end{eqnarray}
where
\begin{eqnarray}
f_1(m,q)&=& -\log 2-\beta^2(1-q^2)/4+(m/2) \mathrm{tanh}^{-1} m \nonumber \\
&+&(1/2) \log(1-m^2),
\end{eqnarray}
and $G_i$ is given by Eq. (\ref{tap}) while $\mathbf{A}$ is the inverse susceptibility matrix. The delta function $\delta (Nq-\sum_i m_i^2)$ enforces the condition that $Q=q$ and ensures that $N_s(f)$ is the number of minima (or saddle-points). The expression $f_1(q,m)$, the single site expression for the free energy, is obtained by using $G_i=0$ to eliminate $J_{ij}$ from Eq. (\ref{FTAP}).

In Ref. \cite{Bray_1980,Bray:81} the details of how one proceeds from Eq. (\ref{Nsdef}) were given in detail and will not be repeated here. One
obtains the following expression for the complexity $\Sigma(f)$ for $f \ge f_c$
 \begin{eqnarray}
\Sigma(f)= \frac{1}{N} \ln \langle N_s(f)\rangle_J &=& -\lambda q -uf -(B+\Delta)(1-q)\nonumber \\&+&(B^2-\Delta^2)/2 \beta^2+\ln I,
\label{Sigmadef}
 \end{eqnarray}
where $I$ is defined by the integral
\begin{eqnarray}
I=\int_{-1}^{1} \frac{dm}{\sqrt{2 \pi P}}\big( \frac{1}{1-m^2}&+&B\big) \exp\bigg[\lambda m^2 +u f_1(m)\nonumber \\&-&\frac{(\tanh^{-1}m -\Delta m)^2}{2 P}\bigg],
\label{Idef}
\end{eqnarray}
where $P= \beta^2 q$. For $f<f_c$ the "annealed" average used in  Eq. (\ref{Sigmadef}) is no longer valid and one must calculate $\langle \ln N_s(f) \rangle_J$ (which is proportional to $N$), and this average  then gives results relevant to a \textit{typical} system. Its determination requires the introduction of replicas to handle the averaging over the logarithm. It turns out also that  full replica symmetry breaking \cite{Bray:81,bray:84} is needed so the calculations become very heavy. Here we shall focus on the case when  $f \ge f_c$ when it is possible to use the annealed average.

The parameters $q, \Delta, \lambda, u, B$ are determined from the stationarity equations for $\Sigma(f)$ for given $f$.
These are
\begin{equation}
\partial \Sigma/\partial \lambda \Rightarrow q =\langle  m^2\rangle.
\label{lamstat}
\end{equation}
Here $\langle \cdots\rangle$ denotes averaging over the weight function of Eq. (\ref{Idef}). The variable $u$ allows us to select the TAP solutions with free energy per spin $f$. Its stationarity equation is
\begin{equation}
\partial \Sigma /\partial u \Rightarrow  f=\langle f_1 \rangle.
\label{ustat}
\end{equation}
Continuing,
\begin{equation}
\partial \Sigma/\partial B\Rightarrow B\big[1-\beta^2 \langle \frac{(1-m^2)^2}{1+B (1-m^2)}\rangle\big]=0,
\label{Bstat}
\end{equation}
\begin{equation}
\partial \Sigma/\partial \Delta \Rightarrow \Delta=-\frac{\beta^2}{2}(1-q)+\langle m \tanh^{-1}m\rangle/(2q),
\label{Deltastat}
\end{equation}
and
\begin{equation}
\partial \Sigma/\partial q \Rightarrow \lambda=B +\Delta-\frac{1}{2q}+ \frac{\langle (\tanh^{-1} m-\Delta m)^2\rangle}{2 \beta^2 q^2} +u \beta^2 q/2.
\label{qstat}
\end{equation}

Solving these equations one finds $B= 0$. The values of $q$, $\lambda$, $\Delta$, and $u$ have to be determined by numerical methods. The complexity is at its largest at $u= 0$. These equations are valid provided
\begin{equation}
x_p =1-\beta^2 \langle (1-m^2)^2\rangle \ge 0
\label{stability}
\end{equation}
As $f\to f_c$, it is found that $x_p$ calculated from these equations goes to zero, indicating that their validity will cease for $f < f_c$.

In Appendix \ref{norm} the "normalization"
$\sum_i v_i^2/N$ and in Appendix \ref{ccalc} the coefficient $c$ of Eq. (\ref{3deriv}) are calculated for the region $f> f_c$ using methods which  are essentially just extensions of those used to obtain the complexity.

\section{Calculation of $\mathbf{2 \beta^2 H}$}
\label{2beta2H}
In this section we shall show that in the large $N$ limit that $2 \beta^2 H = 1$. This is the reason why there is a "null" eigenvalue of the $\mathbf{A}$ matrix and why the coefficient $a$ is zero in the thermodynamic limit. Our demonstration of this is valid for the bond-average over all solutions of free energy $f$, if $f \ge f_c$, (although we would argue that $a= 0$ also for $f  < f_c$).  As in Ref. \cite{aspelmeier:04}, we start by imagining inserting into Eq. (\ref{Nsdef}) the identity
\begin{equation}
1 = \frac{1}{\sqrt{\rm{det} X}}\int_{-\infty}^{\infty}\prod_i\frac{d \phi_i}{\sqrt{2 \pi}} \exp(-\frac{1}{2} \sum_{i,j} \phi_i (X^{-1})_{ij} \phi_j),
\label{iden1}
\end{equation}
This identity which holds provided the matrix $\mathbf{X}$ is positive definite.
We shall set
\begin{equation}
(X^{-1})_{ij}=a_i \delta_{ij}- \beta J_{ij},
\label{adef}
\end{equation}
where
\begin{equation}
a_i = \frac{1}{1-m_i^2} +\beta^2 (1-q).
\label{aidef}
\end{equation}
The definition of $H$ is
\begin{equation}
H=\frac{1}{N}\sum_{i,j}m_i X_{ij} m_j=\frac{1}{N} \sum_{i,j}\langle m_i \langle\phi_i \phi_j\rangle_{\phi}m_j\rangle_{m,J}.
\label{Hphi}
\end{equation}
The average $\langle \cdots \rangle_{m,J}$ is the average over the $m_i$ and the bonds $J_{ij}$.
In order to calculate $H$ we introduce a "field" $ \lambda_0$ and study
\begin{equation}
Z(\lambda_0) =\int W \exp(\beta \lambda_0 \sum_i \phi_i m_i).
\end{equation}
Then
\begin{equation}
\frac{\beta^2}{N}\sum_{i,j} \langle \phi_i m_i \phi_j m_j\rangle = \beta^2 H=\frac{1}{N}\partial^2 \ln Z(\lambda_0)/\partial \lambda_0^2,
\end{equation}
as $\lambda_0 \to 0$. $\langle \cdots \rangle$ is calculated here with the weight function $W$. Note that $\langle \phi_i m_i \rangle = 0$, and $ \partial \ln Z(\lambda_0)/\partial \lambda_0 =0$ as $\lambda_0\to 0$. The bond average is
\begin{equation}
\int \prod_{(ij)} d J_{ij} P(J_{ij})\cdots,
\label{bondavdef}
\end{equation}

The delta functions of $G_i$ can be represented in terms of integrals over $x_i$, which run from $-i\infty$ to $i\infty$. The terms involving
$J_{ij}$ are of the form
\begin{multline}
\int_{-\infty}^{\infty}\prod_{<ij>}dJ_{ij}(N/2\pi )^{1/2}\exp\bigg[
-N\sum_{<ij>}J_{ij}^2/2  \\- \beta \sum_{<ij>}J_{ij}(x_i m_j+x_j m_i+\phi_i \phi_j)\bigg]  \frac{\rm{det}\mathbf{A}}{\sqrt{\rm{det}\mathbf{X}}}.
\end{multline}
so
\begin{multline}
W \sim \int_{-1}^{1}\prod_i dm_i \int \prod_i dx_i \exp\bigg[-\frac{1}{2}\sum_i a(m_i) \phi_i^2  \\+\beta \sum_{(ij)} J_{ij} \phi_i \phi_j-\beta \sum_{(ij)} J_{ij} (x_i m_j+ x_j m_i)\\+\sum_i g(m_i) x_i+\cdots\bigg].
\end{multline}
Averaging over the bonds $J_{ij}$ one gets
\begin{multline}
W \sim  \int_{-1}^{1}\prod_i dm_i \int \prod_i dx_i \exp\bigg[-\frac{1}{2}\sum_i a(m_i) \phi_i^2 \\+\sum_i g(m_i) x_i+\frac{\beta^2}{2 N} \sum_{(i,j)} (x_i m_j+x_j m_i-\phi_i \phi_j)^2\bigg].
\label{bondav}
\end{multline}
The sum over the pairs $(i,j)$ can be extended to all $i,j$ as the diagonal terms with $i=j$ give a negligible contribution when $N$ is large. Then using $Nq=\sum_i m_i^2$, we get
\begin{multline}
 W \sim \int_{-1}^{1}\prod_i dm_i \int \prod_i dx_i \exp\bigg[-\frac{1}{2}\sum_i a(m_i) \phi_i^2  \\+ \frac{\beta^2 q}{2} \sum_i x_i^2+\frac{\beta^2}{2 N} (\sum_i x_i m_i)^2 +\frac{\beta^2}{4N} (\sum_i \phi_i^2)^2 \\-\frac{\beta^2}{N} (\sum_i \phi_i m_i)^2+\sum_i g(m_i) x_i\bigg].
\end{multline}

The terms in the determinant $\det X \{J_{ij}\}$ are effectively shifted to
$\det X\{ J_{ij}-\frac{\beta}{N} (x_i m_j+x_j m_i -\phi_i \phi_j)\}$, and the translation of the $J_{ij}$ by terms of order $1/N$ in the matrix elements of $X_{ij}$ is negligible, allowing the determinant to be separately averaged. (The vanishing of $B$ is then consistent with this neglect).

The square terms are simplified by the Hubbard-Stratonovich identity
\begin{equation}
\exp(a^2/2)=\int_{-\infty}^{\infty} \frac{d  x}{\sqrt{2 \pi}} \, \exp(-x^2/2 +a x).
\end{equation}

We uncouple the square terms involving $(\sum_i x_i m_i)^2$ as follows.
\begin{multline}
\exp\bigg[\frac{\beta^2}{2 N} (\sum_i x_i m_i)^2\bigg] = \sqrt{\frac{N}{2 \pi}}
\int dV \exp\bigg[-\frac{N V^2}{2}\\+ V \beta \sum_i m_i x_i\bigg].
\end{multline}

The square terms involving $(\sum_i \phi_i)^2$ as follows
\begin{multline}
\exp\bigg[\frac{\beta^2}{4 N} (\sum_i \phi_i^2)^2\bigg] =\sqrt{\frac{N}{\pi}}
\int d \rho\exp\bigg[- N \rho^2 \\+ \rho \beta \sum_i \phi_i^2\bigg].
\end{multline}
The cross-term involving $(\sum_i \phi_i x_i) (\sum_j \phi_j m_j)$ is uncoupled via
\begin{multline}
\exp(-\frac{\beta^2}{N} \sum_i \phi_i x_i\sum_j \phi_j m_j)=\frac{N}{ \pi} \int d \eta\, d\eta^* \exp \bigg[-N \eta \eta^*\\+i \beta \eta \sum_i \phi_i x_i +i \beta \eta^* \sum_i \phi_i m_i\bigg].
\end{multline}

The integrals over $V$, $\rho$ and $R$ (see \cite{Bray_1980}), are done by steepest descents. Set $V=-\beta(1-q)-\Delta/\beta$, and $2 R = \beta(1-q) - B/\beta$ where $ \rm{det} \mathbf{X}^{-1}=\prod_i (a_i -2 \beta R) \exp(2 N R^2)$. Similarly $2 \rho=\beta(1-q)-\tilde{B}/\beta$. (We expect $B=0$, $\tilde{B}=B$ as $\lambda_0 \to 0$ when $B$ and $\tilde{B}$ satisfy the same equations). Then doing the $x_i$ integrals (which are up the imaginary axis)
\begin{multline}
Z(\lambda_0) \sim   \int \prod_i d \phi_i \exp\bigg[\beta \lambda_0 \sum_i \phi_i m_i -N \eta \eta^*\\-\frac{1}{2}\sum_i \tilde{a}(m_i)\phi_i^2 +i \sum_i \beta \eta^* \phi_i  m_i\\ -\frac{1}{2 \beta^2 q} \sum_i (\tilde{g}(m_i)+i \beta \eta \phi_i)^2\bigg].
\end{multline}
Note
\begin{multline}
\tilde{g}(m_i)=\tanh^{-1} m_i +\beta^2 (1-q) m_i+ \beta V m_i \\ \to \tanh^{-1}m_i-\Delta m_i,
\end{multline}
while
\begin{equation}
\tilde{a}(m_i)=\frac{1}{1-m_i^2}+B.
\end{equation}
Doing the integrals over $\phi_i$ one gets
\begin{multline}
Z(\lambda_0) \sim \int \prod_i dm_i \exp\bigg[-\frac{1}{2} \sum_i \log   \big[\frac{\tilde{a}(m_i)-\eta^2/q}{\tilde{a}(m_i)}\big]\\-N \eta \eta^{*} -\frac{1}{2 \beta^2 q}\sum_i \tilde{g}(m_i)^2 \\
+\sum_i\frac{(i \beta \eta^{*}m_i -i (\eta/\beta q) \tilde{g}(m_i) +\beta \lambda_0 m_i)^2 }{2 (\tilde{a}(m_i)-\eta^2/q)}\bigg].
\end{multline}
Set $Z(\lambda_0) = \exp(\mathcal{N}/(N \beta^2)$, and $\eta=\beta^2 q \tilde{\eta}$. Note that $\tilde{\eta}$  and $\eta^*$ are of order $\lambda_0$. Then to order $\lambda_0^2$,
\begin{multline}
\mathcal{N}/(N\beta^2)= -A_3 {\eta{^*}}^2/2- \tilde{\eta}^2(A_2-\beta^2 q(1-q))/2 \\+\tilde{\eta} \eta^*(A_1-q) +i \eta^* \lambda_0 A_3 -i\tilde{\eta} \lambda_0 A_1 + \lambda_0^2 A_3/2.
\end{multline}
The coefficients are as in \cite{aspelmeier:04}.
\begin{equation}
A_1=\langle (1-m^2)m (\tanh^{-1} m-\Delta m)\rangle.
\end{equation}
\begin{equation}
A_2=\langle (1-m^2)(\tanh^{-1} m-\Delta m)^2 \rangle.
\end{equation}
\begin{equation}
A_3 =\langle m^2 (1-m^2)\rangle.
\end{equation}
We find useful the identity
\begin{multline}
Max_{x,y}\big[-ax^2/2-by^2/2+cxy+dx +ey\big]\\= \frac{bd^2 + ae^2 +2 cde}{2 (ab-c^2)}.
\label{identquad}
\end{multline}
The maximum occurs at
\begin{equation}
x= \frac{bd+ce}{ab-c^2},
\end{equation}
and
\begin{equation}
y= \frac{ae+cd}{ab-c^2}.
\end{equation}
Then
\begin{equation}
\mathcal{N}/(N\beta^2)=\frac{\lambda_0^2 A_3 q^2}{2((q-A_1)^2+A_3(\beta^2q(1-q)-A_2)},
\end{equation}
so
\begin{equation}
2 \beta^2 H=\frac{2 \beta^2 A_3 q^2}{(q-A_1)^2+A_3(\beta^2q(1-q)-A_2)}.
\label{2beta2Heq}
\end{equation}
This can be shown to equal $1$ at the stationary point by use of the
argument sketched below.

In Ref. \cite{aspelmeier:04} it was demonstrated that $1=2 \beta^2 H$ by solving the stationarity equations numerically to obtain the quantities in Eq. (\ref{2beta2Heq}). Here we shall show that it follows directly from the stationarity equations Eqs. (\ref{lamstat}) - ({\ref{qstat}). With $B= 0$, Eq. (\ref{Idef}) is
\begin{multline}
I =\int_{-1}^{1}\,\frac{dm}{\sqrt{2 \pi q} \beta} \frac{1}{1-m^2}\exp\bigg[\lambda m^2+u f_1(m)\\-\frac{(\tanh^{-1} m -\Delta m)^2}{2 \beta^2 q}\bigg].
\end{multline}
On integrating by parts we get
\begin{multline}
I=-\int_{-1}^{1}\,\frac{dm}{\sqrt{2 \pi q}\beta}\tanh^{-1}m \bigg(2 \lambda m +u f_1^{\prime}(m)\\-\frac{1}{\beta^2 q}(\tanh^{-1}m -\Delta m)\big(\frac{1}{1-m^2}-\Delta\big)\bigg)\\
\times  \exp\bigg(\lambda m^2+u f_1(m)-\frac{(\tanh^{-1} m -\Delta m)^2}{2 \beta^2 q}\bigg).
\end{multline}
This can be re-written as
\begin{multline}
-1= 2\lambda\langle m(1-m^2) \tanh^{-1}m\rangle\\+\frac{u}{2}\langle (1-m^2)(\tanh^{-1}m)^2\rangle -\frac{u}{2}\langle m \tanh^{-1}m\rangle
\\-\frac{1-\Delta}{\beta^2 q}\big(\langle(\tanh^{-1}m)^2\rangle -\Delta \langle m \tanh^{-1}m\rangle\big)\\-\frac{\Delta}{\beta^2 q}\langle m^2 \tanh^{-1}m (\tanh^{-1}m-\Delta m)\rangle.
\end{multline}
This can be put in terms of the coefficients $A_1$, $A_2$, and $A_3$ and with the help of the saddle-point equations themselves simplifies to the relation
\begin{multline}
0=\big(2 \lambda +\Delta u+\frac{\Delta^2}{\beta^2 q}\big)(A_1-q)\\+(\frac{u}{2}+ \frac{\Delta}{\beta^2 q})\big(A_2-\beta^2 q(1-3 q)\big)+\Delta\big(2 \lambda+\frac{\Delta u}{2}\big) A_3.
\label{id1}
\end{multline}
Similarly, integration by parts gives
\begin{multline}
I\langle1-m^2\rangle =\int_{-1}^{1}\frac{dm}{\sqrt{2 \pi \beta^2 q}}\exp\bigg(\lambda m^2+u f_1(m)\\-\frac{ (\tanh^{-1} m -\Delta m)^2}{2 \beta^2 q}\bigg)=-\int_{-1}^{1}\frac{dm}{\sqrt{2 \pi \beta^2 q}}\\m\bigg(2 \lambda m +u f_1^{\prime}(m)-\frac{\tanh^{-1}m-\Delta m}{\beta^2 q} \big(\frac{1}{1-m^2} -\Delta\big)\bigg)\times\\
\exp\bigg(\lambda m^2+ u f_1(m)-\frac{(\tanh^{-1}m-\Delta m)^2}{2 \beta^2 q}\bigg).
\end{multline}
This can be reduced with the help of the stationarity equations to
\begin{multline}
0= \big(2 \lambda+\frac{\Delta u}{2}\big) A_3 +\big(\frac{u}{2}+\frac{\Delta}{\beta^2 q}\big)(A_1-q).
\label{id2}
\end{multline}
Then using Eqs. (\ref{id1}) and (\ref{id2}) the right-hand side of Eq. (\ref{2beta2Heq}) can be shown to equal unity.

\section{Calculation of the normalization $N_z=\sum_i v_i^2/N$}
\label{norm}
Recall that $v_i$ is defines as
\begin{equation}
v_i =\partial m_i /\partial q=\beta^2 \sum_j X_{ij} m_j,
\end{equation}
in this Appendix we shall obtain its \lq\lq normalization"
\begin{equation}
N_z=\frac{1}{N}\sum_i v_i^2= \frac{\beta^4}{N} \sum_i \sum_j \sum_k X_{ij}m_j X_{ik}m_k.
\end{equation}
This quantity plays an important role in our calculations of $\lambda_{min}$ in Eq. (\ref{bound}) and also of $c$ in Appendix \ref{ccalc}.

We will make use of the identity
\begin{multline}
1=\frac{1}{{\rm det}{\mathbf{X}}}\int_{-\infty}^{\infty}\prod_i \frac{d\,\phi_i}{\sqrt{2 \pi}} \prod_i \frac{d\,\rho_i}{\sqrt{2 \pi}}
\\\exp\bigg[-\frac{1}{2}\sum_{i,j} \phi_i (X^{-1})_{ij}\phi_j-\frac{1}{2}\sum_{i,j} \rho_i (X^{-1})_{ij}\rho_j\bigg],
\end{multline}
to write
\begin{equation}
N_z=\frac{\beta^4}{N}\sum_i \sum_j \sum_k \langle \phi_i \phi_j\rangle m_j \langle \rho_i \rho_k \rangle m_k.
\end{equation}
We proceed now as with the calculation of $2 \beta^2 H$. After bond averaging there is now a term (see Eq. (\ref{bondav})),
\begin{multline}
\exp \bigg[ \frac{\beta^2}{4 N}\sum_{i,j}(x_i m_j+x_j m_i -\phi_i \phi_j-\rho_i \rho_j)^2\bigg]=\\
\exp  \bigg[ \frac{\beta^2 q}{2}\sum_i x_i^2 +\frac{\beta^2}{2N}( \sum_i x_i m_i)^2  \\ +
\frac{\beta^2}{4 N} (\sum_i \phi_i^2)^2+ \frac{\beta^2}{4 N} (\sum_i \rho_i^2)^2-\frac{\beta^2}{N}\sum_i \phi_i x_i\sum_j \phi_j m_j\\-\frac{\beta^2}{N}\sum_i\rho_i x_i \sum_j \rho_j m_j+ \frac{\beta^2}{2N} (\sum_i \phi_i \rho_i)^2\bigg].
\end{multline}
We shall introduce as before the term involving $V$ to uncouple the $(\sum_i x_i m_i)^2 $ term, $\eta_1$ and $\eta_1^*$ to uncouple the term $\sum_i \phi_i x_i \sum_j \phi_j m_j$, and $\eta_2$ and $\eta_2^*$ to uncouple the term $\sum_i \rho_i x_i \sum_j \rho_j m_j$.  The term can be re-written using
\begin{multline}
\exp\bigg[\frac{\beta^2}{2N}(\sum_i \phi_i\rho_i)^2 \bigg]\\=\sqrt{\frac{N}{2\pi}}\int_{-\infty}^{\infty} d\,K \exp\bigg[ -\frac{N K^2}{2}+K \beta \sum_i \phi_i \rho_i\bigg].
\end{multline}
Like in Eqs. (B13-B15) we introduce three fields and compute
\begin{multline}
Z(\lambda_1,\lambda_2, \lambda_3)=\int W \exp\bigg [\beta \lambda_1 \sum_i \phi_i m_i\\+ \beta \lambda_2 \sum_i\rho_i m_i +\beta \lambda_3 \sum_i \phi_i \rho_i\bigg].
\end{multline}
Then
\begin{equation}
N_Z=\frac{\partial^3 \ln Z}{\partial \lambda_1\partial\lambda_2\partial \lambda_3},
\end{equation}
in the limit when these fields go to zero. On doing the $x_i$ integrals (which are up the imaginary axis) one gets
\begin{multline}
Z(\lambda_1, \lambda_2,\lambda_3) \sim \int  \prod_i d\, \rho_i d\,\phi_i\frac{1}{{\rm det}\mathbf{X}}\exp\bigg[ \beta \lambda_1 \sum_i \phi_i m_i\\ +\beta \lambda_2 \sum_i \rho_i m_i +\beta \lambda_3 \sum_i \phi_i \rho_i
- N\eta_1\eta_i^{*} +i \beta \eta_1^{*} \sum_i \phi_i m_i \\-N \eta_2 \eta_2^{*} +i \beta \eta_2^{*} \sum_i \rho_i m_i-NK^2/2+K\beta \sum_i \phi_i\rho_i \\
-\sum_i\big[ \frac{(\tilde{g}(m_i) +i \beta \eta_1 \phi_i +i \beta \eta_2 \rho_i)^2}{2 \beta^2 q}\\-\frac{1}{2} \tilde{a}(m_i) \phi_i^2 -\frac{1}{2} \tilde{a}(m_i) \rho_i^2 \big]\bigg] \\
\sim \int   \prod_i d\,\phi_i d\rho_i \frac{1}{{\rm det}\mathbf{X}} \exp\bigg[ -N \eta_1 \eta_1^{*}- N \eta_2 \eta_2^{*} - NK^2/2
 \\
+ \sum_i\big[\i \beta \eta_1^{*} \phi_i m_i + i \beta \eta_2^{*} \rho_i m_i- \frac{\tilde{g}(m_i)^2 }{2 \beta^2 q}\\-i  \eta_1 \tilde{g}(m_i) \phi_i/(\beta q)-i  \eta_2\tilde{g}(m_i) \rho_i/(\beta q) \\+\eta_1 \eta_2 \phi_i\rho_i/q +\beta \lambda_1 \phi_i m_i +\beta \lambda_2 \rho_i m_i +\beta \lambda_3 \phi_i \rho_i\
+\\ K \beta\phi_i \rho_i-\frac{1}{2} (\tilde{a}(m_i)-\eta_1^2/q) \phi_i^2-\frac{1}{2} (\tilde{a}(m_i)-\eta_2^2/q)\rho_i^2 \big]\bigg].
\end{multline}

We next use the identity of Eq. (\ref{identquad}) to do the integrals over $\phi_i$ and $\rho_i$.
Set $a=\tilde{a}(m_i) -\eta_1^2/q$, $b=\tilde{a}(m_i)-\eta_2^2/q$, $c=\beta(K +\lambda_3 +\eta_1 \eta_2 /(\beta q))$, $d= \beta \lambda_1 m_i -i \eta_1 \tilde{g}(m_i)/(\beta q) +i \beta \eta_1^*m_i$, $e= \beta \lambda_2 m_i -i \eta_2 \tilde{g}(m_i)/(\beta q) + i \beta \eta_2^{*} m_i$. Then

\begin{multline}
Z\sim \int W\prod_i \,d\,m_i \exp\bigg[- NK^2/2- N \eta_1 \eta_1^* -N \eta_2 \eta_2^*\\+\sum_i\big[\frac{bd^2+ae^2+2 cde}{2(ab-c^2)}-\frac{\tilde{g}(m_i)^2}{2\beta^2q}-\frac{1}{2}\log\big[\frac{a b-c^2}{\tilde{a}(m_i)^2} \big] \bigg].
\end{multline}

Introduce $ \tilde{K}=K + \lambda_3 +\eta_1 \eta_2 /(\beta q)$. To quadratic order the argument of the exponential is

\begin{multline}
Arg = -\frac{N}{2}[\tilde{K}-\lambda_3-\eta_1\eta_2/(\beta q)]^2 -N \eta_1 \eta_1^* -N \eta_2 \eta_2^*+\sum_i\\
\big[-\frac{1}{2 \beta^2 q}  \tilde{g}(m_i)^2
+ \big[\frac{bd^2+ae^2+2\beta \tilde{K}de}{2(ab-\beta^2 \tilde{K}^2)} \\-\frac{1}{2} \log\frac{\tilde{a}(m_i)^2 -\beta^2 \tilde{K}^2}{\tilde{a}(m_i)^2}\big].
\end{multline}

We now eliminate $\eta_1,\eta_1^*,\eta_2,\eta_2^*$. These are of order of the $\lambda_i$.

We will take it that in the limits of $\lambda_1, \lambda_2, \lambda_3 \to 0$, then $\eta_1, \eta_1^* \sim \lambda_1$ and $ \eta_2, \eta_2^*\sim \lambda_2$.
The stationarity equation for $\tilde{K}$ then is
\begin{equation}
-N(\tilde{K}-\tilde{\lambda}_3)+ \beta^2 \tilde{K}\sum_i (1-m_i^2)^2/(1-\beta^2 \tilde{K}^2 (1-m_i^2)^2=0,
\end{equation}
where terms of higher order in the $\lambda_i$ have been dropped. Note that here $\tilde{\lambda}_3= \lambda_3 +\eta_1\eta_2/(\beta q)$.  We shall now work close to the critical value of $u_c$ where
\begin{equation}
\tau \equiv x_p=1-\beta^2 \frac{1}{N} \sum_i (1-m_i^2)^2
\end{equation}
is small. We shall use the notation $\tau$ for $x_p$ when it is small and to emphasize its similarity with the variable $(1-T/T_c)$ in critical behavior phenomena. Then the equation for $\tilde{K}$ reduces to
\begin{equation}
\tilde{\lambda}_3= \tau \tilde{K}-c_4\tilde{K}^3 +\cdots,
\end{equation}
where $c_4=\beta^2 \frac{1}{N}\sum_i(1-m_i^2)^4$. Its solution is of the form
\begin{equation}
\tilde{K}= \frac{\tilde{\lambda}_3}{\tau} F(c_4 \tilde{\lambda}_3^2/\tau^3).
\end{equation}
The function $F(x)$ goes to $1$ as $x\to 0$ and goes as $1/\sqrt{x}$ as $x\to \infty$. We shall work in the limit of small $x\equiv c_4 \tilde{\lambda}_3^2/\tau^3$.

The terms in Eq. (C14) give a contribution to Arg
\begin{equation}
Arg/N= -\tau \tilde{K}^2/2 +\tilde{K}\tilde{\lambda}_3 +c_4 \tilde{K}^4/4,
\end{equation}
Then the leading contribution at small $x$, where $\tilde{K}=\tilde{\lambda}_3/\tau$, is
\begin{equation}
Arg/N=\tilde{\lambda}_3^2/(2 \tau).
\end{equation}
This gives a contribution to Arg of
\begin{equation}
Arg=-N \frac{\lambda_1 \lambda_2\lambda_3}{4 \tau\beta q},
\end{equation}
using $\eta_1=i\lambda_1/2$ and $\eta_2=i \lambda_2/2$ (see below).

Put $\eta_1 =\beta^2 q \tilde{\eta}_1$, and $\eta_2 =\beta^2 q \tilde{\eta}_2$. We want the term of order $\lambda_1\lambda_2\lambda_3$, in the limit when all the $\lambda_i\to 0$. We thus need to determine the dependence of $\eta_1, \eta_1^*, \eta_2, \eta_2^*$ on $\lambda_1, \lambda_2$ in the contribution to Arg from

\begin{multline}
Arg \sim  -N \eta_1 \eta_1^* -N \eta_2 \eta_2^* \\
+ \big[\tilde{a}(m_i)(i\beta \eta_2^*m_i-i \eta_2 \frac{\tilde{g}(m_i)}{\beta q} +\beta \lambda_2 m_i)^2 +\frac{\eta_2^2 \tilde{a}(m_i)}{2q}\\ +
\tilde{a}(m_i)(i\beta \eta_1^*m_i-i \eta_1 \frac{\tilde{g}(m_i)}{\beta q} +\beta \lambda_1 m_i)^2+\frac{\eta_1^2 \tilde{a}(m_i)}{2q} + 2 \beta \tilde{K}\times \\
(i\beta \eta_2^*m_i-i \eta_2 \frac{\tilde{g}(m_i)}{\beta q} +\beta \lambda_2 m_i)(i\beta \eta_1^*m_i-i \eta_1 \frac{\tilde{g}(m_i)}{\beta q} +\beta \lambda_1 m_i)\big] \\
  /(2({\tilde{a}(m_i)^2 -\beta^2 \tilde{K}^2)}.
\end{multline}

This expression can be handled using the "quadratic" formulae, Eq. (\ref{identquad}) for  maximization first in the $\eta_1, \eta_1^*$ sector, then in the $\eta_2, \eta_2^*$ sector to get an expression involving $\lambda_1,\lambda_2$ and $\tilde{K}_0$. We then pick out the term in $\lambda_1\lambda_2 \lambda_3$.

The terms in $\eta_1$ and $\eta_2$ are decoupled as the coupling terms are small: the term in the $1$ variables in Arg is
\begin{multline}
Arg/(N \beta^2)= -q\tilde{\eta_1} \eta_1^{*}+(1/2N)\sum_i\\\big[(1-m_i^2)(i \eta_1^*m_i-i \tilde{\eta}_1\tilde{g}(m_i)+\lambda_1 m_i)^2 +\beta^2 q (1-m_i^2) \tilde{\eta}_1^2\big]\\
=(A_1-q) \tilde{\eta}_1 \eta_1^{*}-(1/2) [A_2-\beta^2 q(1-q)]\tilde{\eta}_1^2 \\-(1/2) A_3 (\eta_1^{*})^2+i \lambda_1A_3 \eta_1^{*}-i \lambda_1 A_1\tilde{\eta}_1 +(1/2) A_3 \lambda_1^2.
\end{multline}
 (Compare with Eq. (B20)). Note that we do not need the changes to $\eta_1$ which are of order $K$ from including the cross term in 1 and 2 terms Eq. (C19) as such terms modify Arg in Eq. (C19) at order $K^2$. Eq. (C20) is evaluated at the stationary point and so changes to the values of $\eta_1$ and $\eta_1^*$ of order $K$ change that expression for Arg to order $K^2$.  The maximum occurs when $\eta_1= i \lambda_1/2$ and $\eta_2=i\lambda_2/2$ on using Eqs. (B25) and (B26).
 \begin{multline}
\tilde{\eta}_1=\frac{i q A_3 \lambda_1}{(q-A_1)^2+A_3(\beta^2 q(1-q)-A_2)}\\= \frac{i q A_3}{ 2 \beta^2 A_3 q^2}= \frac{i\lambda_1}{2 \beta^2 q}.
\end{multline}
Then
 \begin{equation}
 \eta_1^*-i \lambda_1= i \lambda_1 \frac{A_1-q}{2 \beta^2 q A_3}
 \end{equation}
 and
 \begin{equation}
 \eta_2^*-i \lambda_2= i \lambda_2  \frac{A_1-q}{2 \beta^2 q A_3}
 \end{equation}

Define $dv_i(\lambda_1)= \beta \lambda_1 m_i -i \eta_1 \tilde{g}(m_i)/(\beta q) +i \beta \eta_1^*m_i$ then becomes
\begin{equation}
dv_i(\lambda_1)=\frac{\lambda_1}{2  q}\bigg(\tilde{g}(m_i)-  m_i\frac{A_1-q}{ A_3}\bigg)(1-m_i^2).
\end{equation}
Note that
\begin{equation}
\frac{1}{N} \sum_i m_i dv(\lambda_1)= \lambda_1/2.
\end{equation}
Without the factor $\lambda_1$, $dv_i$ is essentially $v_i$.

The coefficient of the crossterm in $\tilde{K}$ in Eq. (C19) can be written as
\begin{equation}
{\rm cross}\,\, \lambda_1 \lambda_2= \beta \sum_i dv_i(\lambda_1) dv_i(\lambda_2)/N.
 \end{equation}
and together with the other term in $\lambda_1\lambda_2\lambda_3$ in Eq. (C18) for $\lambda_1\lambda_2 \lambda_3$ we get
\begin{eqnarray}
\frac{1}{N}\sum_i v_i^2=\frac{{\rm cross}\,(1-\tau) -1/(4 \beta q)}{\tau}
\end{eqnarray}
which reduces as $f \to f_c$ to
\begin{equation}
\frac{1}{N}\sum_i v_i^2=\frac{0.813 241}{\tau}.
\end{equation}
Thus the normalization term $N_z$ diverges as $\sim 1/\tau$ as $\tau \to 0$.
\section{The cubic term C = (1/N)$\sum_i 2 m_i v_i^3/(1-m_i^2)^2$}
\label{ccalc}
As before,
\begin{equation}
v_i =\beta^2\sum_j X_{ij} m_j.
\end{equation}
Note that $c= C-3 \beta^2 \sum_i v_i^2/N$, according to Eq. (\ref{3deriv}).
We shall write
\begin{equation}
C= \frac{1}{N} \sum_i \sum_{j,k,l} \frac{2 \beta^6 m_i}{(1-m_i^2)^2}\langle\phi_i \phi_j\rangle m_j \langle \rho_i \rho_k\rangle m_k\langle \tau_i \tau_l\rangle m_l.
\end{equation}
Then (see Eq.(B10))
\begin{multline}
\exp \bigg[ \frac{\beta^2}{4 N}\sum_{i,j}(x_i m_j+x_j m_i -\phi_i \phi_j-\rho_i \rho_j-\tau_i \tau_j)^2\bigg]=\\
\exp  \bigg[ \frac{\beta^2 q}{2}\sum_i x_i^2 +\frac{\beta^2}{2N}( \sum_i x_i m_i)^2  \\+
\frac{\beta^2}{4 N} (\sum_i \phi_i^2)^2+ \frac{\beta^2}{4 N} (\sum_i \rho_i^2)^2+\frac{\beta^2}{4 N} (\sum_i \tau_i^2)^2 \\-\frac{\beta^2}{N}\sum_i \phi_i x_i\sum_j \phi_j m_j-\frac{\beta^2}{N}\sum_i\rho_i x_i \sum_j \rho_j m_j\\-\frac{\beta^2}{N}\sum_i \tau_i x_i\sum_j \tau_j m_j + \frac{\beta^2}{2N}(\sum_i\rho_i \phi_i)^2\\+ \frac{\beta^2}{2N} (\sum_i \phi_i \tau_i)^2+ \frac{\beta^2}{2N} (\sum_i \rho_i \tau_i)^2 \bigg].
\end{multline}
We introduce similar fields as in Eq. (C7):
\begin{multline}
Z(\lambda_1, \lambda_2,\lambda_3, \lambda_4)=\int W
\exp\bigg[\beta \lambda_1 \sum_i \phi_i m_i\\ +\beta \lambda_2 \sum_i \rho_i m_i
+\beta \lambda_3 \sum_i \tau_i m_i
+ \beta^3  \lambda_4 \sum_i \frac{\phi_i \rho_i \tau_i m_i}{(1-m_i^2)^2}\bigg],
\end{multline}
and calculate
\begin{equation}
C=2 \frac{\partial^4 \ln Z(\lambda_1,\lambda_2,\lambda_3,\lambda_4) }{\partial \lambda_1 \partial \lambda_2\partial\lambda_3\partial\lambda_4},
\end{equation}
in the limit when $\lambda_1, \lambda_2,\lambda_3$ and $\lambda_4 \to 0$. The term in $\lambda_4$ can be handled by pretending it is imaginary and doing the Airy style integral. In practice it is easier to progress by recognizing that in the limit  when $\lambda_1, \lambda_2,\lambda_3$ and $\lambda_4 \to 0$, the coupling between the 1,2, and 3 sectors is small and then one can approximate the term in $\lambda_4$
\begin{equation}
\beta^3 \lambda_4 \frac{\phi_i\rho_i \langle \tau_i \rangle m_i}{(1-m_i^2)^2}.
\end{equation}
There are  two other permutations involving $\langle \rho_i\rangle$ and $\langle \phi_i \rangle$. We  introduce $\eta_1, \eta_1^*$, $\eta_2,\eta_2^*$, and $\eta_3, \eta_3^*$ to uncouple the terms of the form $\sum_i \phi_i x_i \sum_j\phi_j m_j$ and terms $K_1,K_2,K_3$ to uncouple the terms of the form $(\sum_i \rho_i \phi_i)^2$.

The leading order the terms in (say) $K_1$ will be small and the stationarity equation for $K_1$ in terms of $\lambda_4$ is (see Eq. (C17)) is
\begin{equation}
-N K_1+ \sum_i \beta^2 (K_1+\beta^2 \lambda_4 dv(i) \frac{m_i}{(1-m_i^2)^2}) (1-m_i^2)^2.
\end{equation}
Hence
\begin{equation}
K_1= \frac{\beta^2 \lambda_4}{2 \tau}.
\end{equation}
 Then the relation between $\lambda_3$ of the normalization calculation (Eq.(C14)) (and its effective value as calculated from Eq. (D8)  and $K_1$ is
$\lambda_3 \equiv  \beta^2 \lambda_4/ (2 (1-\tau))$. (The factor $(1-\tau)$ arises from the difference between $K$ and $\tilde{K}$.)
The rest of the calculation gives $\bigg(\beta^2 /(2(1- \tau) N )\bigg) \sum_i  v_i^2$.
Hence we can write
\begin{multline}
d^3 F/dq^3 = -3 \beta^2 \sum_i v_i^2 +\sum_i \frac{2 m_i v_i^3}{(1-m_i^2)^2}\\= -3 \beta^2 \sum_i v_i^2(1-1/(1-\tau))= \frac{3 \beta^2 \tau}{1-\tau}\sum_i v_i^2.
\end{multline}
remembering the other two permutations.

With this form $c$ stays finite as we approach $f_c$, where $\tau \to 0$, (as $\sum_i v_i^2/N\sim 1/\tau$). Right at $f_c$, $c=2.439723 \beta^2$ on using Eq. (C28), and is finite for all $f > f_c$.

\bibliography{refs}

\end{document}